\documentclass{aa}
\usepackage{graphicx}
\usepackage{natbib}
\usepackage{times}
\begin{document}

\newcommand{\kms}{km\,s$^{-1}$}
\newcommand{\cmm}{cm$^{-2}$}
\newcommand{\hco}{HCO$^{+}$}
\newcommand{\cch}{C$_{2}$H}
\newcommand{\choh}{CH$_{3}$OH}
\newcommand{\etal}{et al.\ }
\newcommand{\x}{$\times$}

\title{Chemistry in Disks. \\
II. -- Poor molecular content of the AB Aur disk.
\thanks{Based on observations carried out with the IRAM
Plateau de Bure Interferometer.
IRAM is supported by INSU/CNRS (France),
MPG (Germany) and IGN (Spain).}}

\author{
Katharina Schreyer \inst{1},
St\'ephane Guilloteau \inst{2,3},
Dmitry Semenov \inst{4},
Aurore Bacmann \inst{2,3},
Edwige Chapillon \inst{5},  
Anne Dutrey \inst{2,3},
Frederic Gueth \inst{5},
Thomas Henning \inst{4},
Frank Hersant  \inst{2,3}, 
Ralf Launhardt \inst{4},
Jer\^ome Pety \inst{5},
Vincent Pi\'etu \inst{5}
}

\offprints{K. Schreyer,  \email{martin@astro. uni-jena.de}}

\institute{
Astrophysikalisches Institut und Universit\"ats-Sternwarte,
Schillerg\"asschen 2-3 D - 07745 Jena, Germany
\and
Universit\'e Bordeaux 1; Laboratoire d'Astrophysique de Bordeaux (LAB)
\and{} CNRS/INSU - UMR5804 ; BP 89, F-33270 Floirac,
France
\and
Max-Planck-Institut f\"ur Astronomie, K\"onigstuhl 17, D-69117
Heidelberg, Germany
\and
IRAM, 300 rue de la piscine, F-38406 Saint Martin d'H\`eres, France
}
\titlerunning{Molecules in the AB Aur disk}

 \date{Received  20 December 2007 / Accepted  10 September 2008}
\abstract{}
%
{We study the molecular content and chemistry of a circumstellar disk surrounding the Herbig Ae star AB Aur
at (sub-)millimeter wavelengths. Our aim is to reconstruct the chemical history and composition of the AB
Aur disk and to compare it with disks around low-mass, cooler T Tauri stars. }
%
{We observe the AB Aur disk with the IRAM Plateau de Bure Interferometer in the C- and D- configurations
in rotational lines of CS, HCN, C$_2$H, CH$_3$OH, HCO$^+$, and CO isotopes.
Using an iterative minimization technique, observed columns densities and abundances are derived.
These values are further compared with results of an advanced chemical model that is based on a steady-state
flared disk structure with a vertical temperature gradient, and gas-grain chemical network with surface
reactions.
}
%
{We firmly detect HCO$^+$ in the 1--0 transition, tentatively detect HCN, and do not detect CS, C$_2$H,
and CH$_3$OH. The observed HCO$^+$ and $^{13}$CO column densities as well as the upper limits to the column densities of HCN,
CS, C$_2$H, and CH$_3$OH are in good agreement with modeling results and those from previous studies.
}
%
{The AB Aur disk possesses more CO, but is less abundant in other molecular species compared to the DM Tau disk.
This is primarily caused by intense UV irradiation from the central Herbig A0 star, which results in a hotter disk
where CO freeze out does not occur and thus surface formation of complex CO-bearing molecules might be inhibited.}
\authorrunning{Schreyer \etal}
\maketitle

%
\section{Introduction}
The rich variety of the detected exoplanetary systems cannot be fully
understood without knowledge about the physical and chemical
evolution of their precursors -- protoplanetary disks.
The question  what role chemistry  plays during
planet formation and how it is coupled to disk dynamics remains loosly constrained.

Nowadays it is widely believed that disk evolution is controlled by
redistribution of angular momentum due to turbulent viscosity. A promising source
of turbulence is the
magnetorotational instability that works if the disk matter is sufficiently
ionized \citep{BH_91}. The chemistry of ionization fraction in protoplanetary disks
was studied in detail, both theoretically and observationally
\citep[see, e.g.,][]{Gammie_96,GNI_97,IG_99,Red2,Ilg06a,Pascucci_ea07}.
To a large extent it is dominated by stellar radiation in the upper disk region and high-energy
cosmic ray particles in their midplanes. For the inner, planet-forming disk regions, X-ray radiation from a
young star may play a crucial role. Recently, \citet{Tur07} 
investigated the evolution of the ionization degree in the inner region of a protoplanetary disk, using a coupled
3D magneto-hydrodynamical and chemical code and found that dynamic transport increases
electron concentration within the dead zone. It has been widely discussed that
the interplay between chemical and transport processes must have been important during
the formation of our own solar system \citep[e.g.,][]{Cyr98,BoMo02,Woo05}.

\begin{table*}[th]
\caption{List of observations using the Plateau de Bure Interferometer.
In the column of the configuration, the abbreviation of  e.g.\ ``5D''
refers to the number of antennas and the used configuration. }
\label{obspara}
\centering
\begin{tabular}{c c c c c c c c c}
\hline\hline  
Line  & transition &  frequency &configu- & baseline  & integration   &  synthezied beam  & 1 $\sigma$ rms\\
          &                    &     [MHz]       &ration     &  range [m] &  time ($T_{\rm ON}$)  [h]       &   size, position angle &  [mJy/beam] \\
\hline
HCO$^+$     &1--0                                                     & 89188.523 & 5D,C2   &  24 -- 175  & 5.9+10   &    5.2$''$ \x\ 4.8$''$, 6$^\circ$ &  18 \\          
HCN       &  $J$=1$-$0, $F$=2$-$1                        &  88631.847&6CD &  24 -- 175 & 5.9           & 6.0$''$ \x\ 4.2$''$, 105$^\circ$  &  24 \\      
CS                 & 2--1                                                    & 97980.950 & 5D    &  24 -- 160   & 1.8          & 9.0$''$ \x\ 4.2$''$, 120$^\circ$   & 170\\        
C$_2$H & $N$ = 1--0,  $J$ = 3/2--1/2, $F$=2--1 & 87316.925 & 6D    & 24 -- 113 & 1        &  9.8$''$ \x\ 3.9$''$, 110$^\circ$  & 37 \\          
CH$_3$OH &15$_{(3,13)}-$14$_{(4,10)}$ A$^+$    & 88594.960 & 6CD &   24 -- 175 & 3.9         &  6.1$''$ \x\ 4.2$''$, 105$^\circ$  &  24\\       
CH$_3$OH &15$_{(3,12)}-$14$_{(4,11)}$ A$^-$     & 88940.090 & 5D    &   24 -- 175 & 3.9+2.2  &  6.3$''$ \x\ 4.3$''$, 105$^\circ$  &  24\\         
CH$_3$OH &   2$_{(1,1)}-1_{(1,0)} A^-$                    & 97582.830 & 5D    &   32 -- 73   & 1.8          &  9.4$''$ \x\ 4.3$''$, 148$^\circ$   & 170  \\        
\hline
\end{tabular}
\end{table*}   

Due to high computational demands, the studies of disk chemistry coupled to disk dynamics are
in their early stages
\citep{Ilgner_etal2004,Willacy_etal2006,Semenov_etal2006,TG07}.
Chemical models were successfully applied to study disk chemistry for steady-state accretion disks
\citep[e.g.,][]{Will98, vanDishoeck_Blake1998,Aikawa_Herbst1999,Markwick_ea02,Sem05,Dut07}.
The most important result is the chemical stratification in such disks, where many molecules are abundant
at an intermediate, slightly UV irradiated layer, while being frozen out in the cold and dark midplane and
broken apart by unshielded high-energy radiation in disk atmospheres. \citet{Bergin_ea03} have shown
that the non-thermal UV radiation from T Tauri stars may partly come in Ly$_\alpha$ photons, which significantly
affects abundances of molecules dissociated by these photons.
In contrast, Herbig Be/Ae stars emit much stronger thermal UV radiation, with
circumstellar disks being hotter and more ionized than the disks around T Tauri stars.

A detailed understanding of physical and chemical conditions of the planet-forming environment
requires not only sophisticated models, but also high-resolution observations of
protoplanetary disks, both in dust continuum and various molecular lines.
Unfortunately, such studies are challenging due to the limited sensitivity and spatial
resolution of available observational facilities, and small disk sizes ($\sim 100-1\,000$~AU).
\citet{Dut97} and \citet{Kas97} have first detected several molecular species
towards protoplanetary disks by using the IRAM 30m single-dish telescope, followed by observations of
\citet{vZad01,Thi04,Sem05}. Interferometric observations by \citet{Qi01,Aik03,Pie06,Pie07}
allowed to study disk gas at outer radii $\ge 50$~AU in lines of several abundant
species (CO, $^{13}$CO, C$^{18}$O, \hco, CS, CN, H$_2$CO). These studies have confirmed that
photochemistry plays an important role even at large disk radii and that abundances of many gas-phase
molecules are depleted.

In 2005 a joint Heidelberg-Bordeaux ``Chemistry In Disks'' (CID) project was established
to investigate the spatial distribution of various molecular tracers across well-studied T~Tauri and Herbig~Ae
disks of various age, followed by comprehensive modeling. In the first paper of this series, we have presented the results of
a deep search for N$_2$H$^+$ and HCO$^+$ towards two T Tauri stars (DM~Tau, LkCa~15) and one Herbig Ae star
(MWC~480), see \citet{Dut07}. The N$_2$H$^+$ emission has been detected in LkCa 15 and DM Tau, with the N$_2$H$^+$ to HCO$^+$
ratio similar to that of cold dense cores and the disk ionization degree as predicted by chemical models.
In this second paper, we investigate the molecular content of the well-studied disk around the Herbig Ae star
AB Aurigae \citep[][distance = 145 pc]{vdAnc98,Gra99,Rob01,Fuk04}, which is an
intermediate-mass analog of the young T~Tauri stars and evolutionary precursor of the
main-sequence debris disk sources like $\beta$\,Pic, $\alpha$\,Lyr, and $\alpha$\,PsA.
This disk is peculiar in comparison with other known circumstellar disks since it is still embedded in a
rather extended envelope and has clumpy density sub-structures \citep{Fuk04,Cor05,Pie05,Lin06}.
This points toward a low-mass companion or a giant planet \citep{Rod07}, a recent stellar encounter, or
imprint of earlier non-steady evolution (see discussion in \citet{Pie05}, hereafter Paper~II).
Using the IRAM 30-m telescope, \citet[][hereafter Paper I]{Sem05} have studied the molecular content towards
this source at low resolution and derived basic parameters of the envelope.
They have derived a low disk inclination of about $17^{\rm o}$ (face-on orientation) and low disk mass
of about 0.013\,$M_{\sun}$. \citet[][]{Pie05} have
resolved the disk structure in different CO isotopes using the Plateau de Bure Interferometer (PdBI).
They have found that the disk is inclined by about $30^{\rm o}$ and its rotation departs from the Keplerian law,
with an exponent for the rotation velocity as low as 0.41.
The outer radius of the gaseous disk is close to $\sim 1000$~AU, 
its inner hole is at a radius of  $\approx$\,100~AU 
in the dust emission,  and at $\approx$\,70~AU in CO lines, and the disk mass is $\approx$ 0.02\,$M_{\sun}$.

The main aim of the present study is to search for C$_2$H, CS, HCN, HCO$^+$, CH$_3$OH, and CO isotopes
in the disk of AB Aur using compact interferometer configurations of the PdBI, to derive their
column densities, and to compare these values with those from a robust chemical model and the disk of a
cooler T~Tauri star DM~Tau.

\section{Observations}
Observations with the PdBI were carried out
between February and August 2002 using the compact C and D
configurations and single side band tuning.
Table~\ref{obspara} lists the observational parameters for the individual observing runs.

\begin{figure*}[th]
\includegraphics[width=0.58\textwidth,angle=270,clip=1]{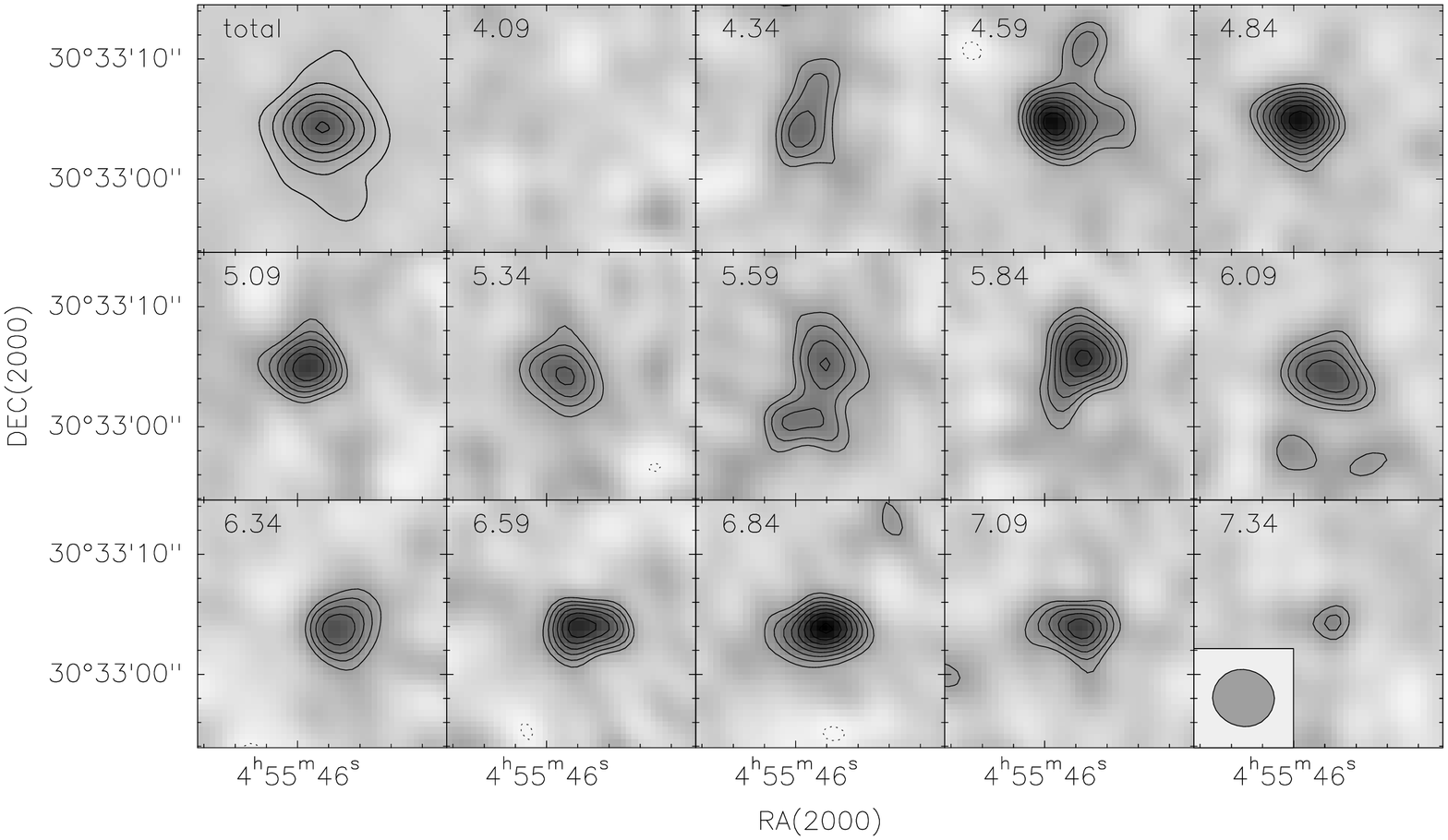}     
\caption{Mapping results of the combined \hco\ (1--0) data.
Contour levels are:      -0.04   to  0.13 Jy \kms\ in step of 0.04 Jy \kms.
The mean velocity of each channel is given in each box top left.
In the panel top left, the total integrated \hco\ map is shown.
Contour levels are:  -0.015 to  7  0.10 Jy \kms\ in step of 0.015 Jy \kms.
The area of the HPBW of the synthezied beam is indicated as a grey ellipse
in the panel bottom right.}
\label{fig1}
\end{figure*}

The frequencies are taken from the Cologne Database for Molecular Spectroscopy
\citep[][2005]{Mue01}.
The phase reference center is
RA(2000) = 04$^{\rm h}$\,55$^{\rm m}$\,54.8$^{\rm s}\!\!,$  DEC(2000) =
+30$^{\circ}$\,33$'$\,04.3$''\!\!.$
All line observations were performed in the lower side band
with a frequency resolution of 
39~kHz, leading to a velocity resolution of 0.14 \kms.
We took advantage of the flexible correlator to observe simultaneously
transitions of \choh\ in the same bandpass as for the observations
of \hco, CS, and HCN. The total integration time was rather short, about 2 hours
for each line. We combined the HCO$^+$ (1--0) data obtained during these observations with that from
Paper~II to reconstruct the final image at 5.34$''$ $\times$ 4.8$''$ spatial resolution.

The bandpass and phase calibration were
performed using observations of the objects 3C84, 3C345, MWC349,
CRL 618, and 0528+134.
The GILDAS software is utilized
for the data reduction and final phase calibration.
The resulting synthesized half-power beam widths are
listed in Table~\ref{obspara}.

All listed lines in Table \ref{obspara} were centered to the systemic
velocity of V$_{\rm LSR} = 5.8$ \kms.
The weak underlying continuum was detected, with an
integrated $\lambda$\,3~mm flux of 3.7~mJy. This is consistent with previous measurements of \citet{Pie05}.
For the final analysis of the observed line spectra this continuum was subtracted.

\section{Results}
\begin{figure*}
 \includegraphics[width=0.63\textwidth,angle=270,clip=1]{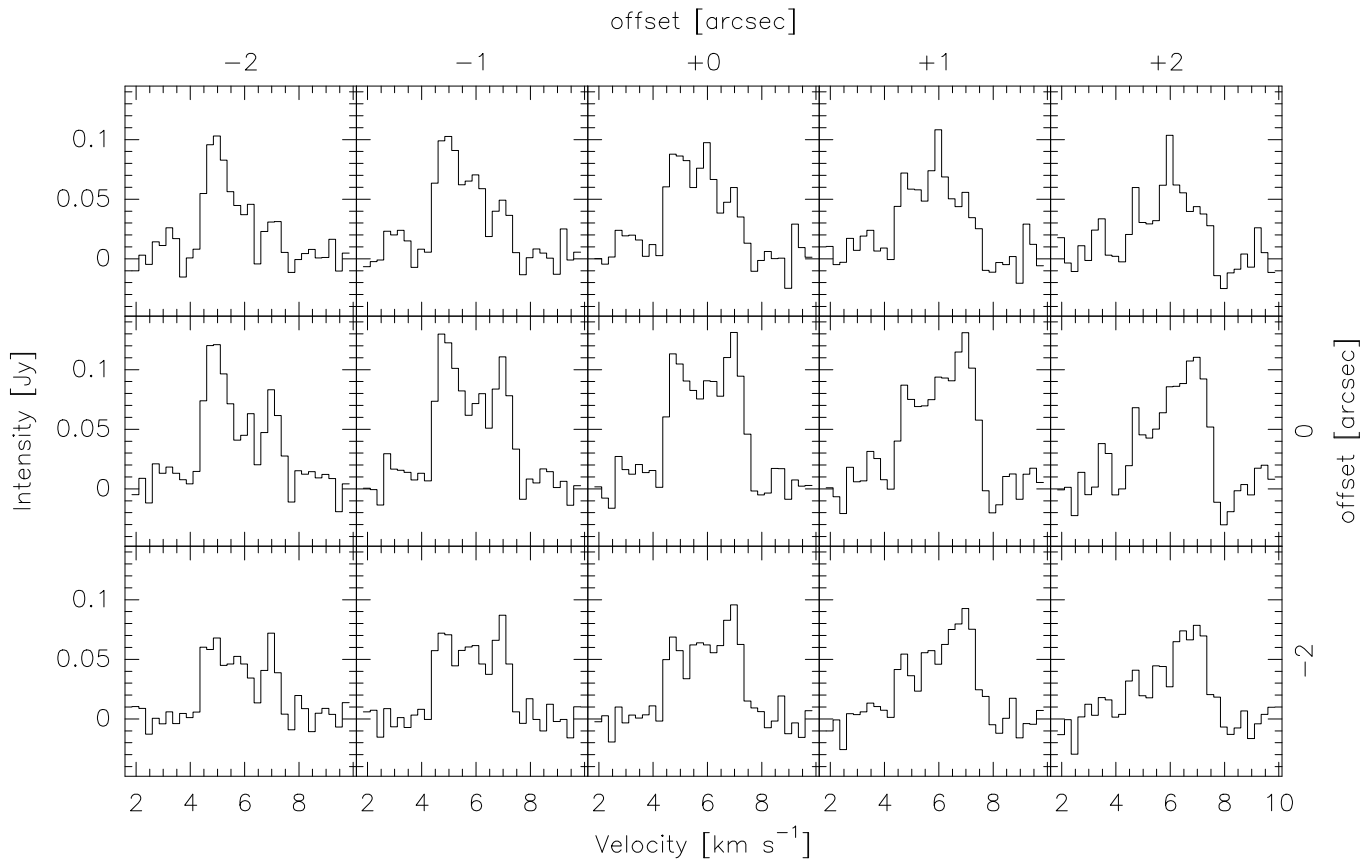}   
\caption{Spatial variation of the HCO$^+$ (1--0) line profiles across the AB Aur
disk.   }
\label{fig2}
\end{figure*}

The transition of \hco\ (1--0) was the only spectral line that
was well detected in each observing run. In Figs.~\ref{fig1}
and \ref{fig2} we show the combined spectrum of \hco\ (1--0).
With the beam of 5.34$''$ $\times$ 4.8$''\!\!,$ that is comparable to the disk size,
we do not fully resolve the disk around AB Aur (Fig.~\ref{fig1}, upper left panel).
However, the pattern typical for rotating circumstellar disks appears clearly in various channel maps (Fig.~\ref{fig1}).
The peak intensity of the double line profile of \hco\
is 0.12~Jy/beam, which is 10 times higher than the noise level.
Fig.~\ref{fig2} shows the variation of the line profiles across the spatial extent of the AB Aur
disk, which is similar to the spectra presented in Papers I/II. The symmetric double-peaked line profile at the
center of the spectral map is a typical feature of barely resolved Keplerian disks. Since deconvolution
is a highly non-linear procedure that is prone to errors we perform our analysis in the $uv$-space, using
the $\chi^2$-minimization technique of \citet{Gui98}.

None of the other attempted species were detected (C$_2$H, CS, and CH$_3$OH),
but we have a hint of the HCN (1--0) hyperfine line components $F$ = 1-1 and $F$ = 2-1
at the 2\,$\sigma$ noise level. The 1\,$\sigma$ rms noise levels for all species
are listed in Table~\ref{obspara}.

To better quantify the column densities and molecular abundances, we analyze the data by
applying the parametric disk model and the $\chi^2$-minimization method as described in \citet{Gui98}.
First, we analyze the HCO$^+$ data leaving as many free parameters as possible.
This allows us to verify the compatibility of the HCO$^+$ emission with the disk parameters derived from
the CO isotopologues in Paper II. The disk parameters derived from Paper II and found from
HCO$^+$ are presented in Table~\ref{initial_params}.

Although the angular resolution is low, the HCO$^+$ data confirms one important point: the existence of the inner
hole of $\simeq 75-100$ AU radius found in Paper II. This hole manifests itself as a lack of emission at projected
velocities beyond $\pm 1.6$ km\,s$^{-1}$ from the systemic velocity, but the hole is not directly visible
in the spectral map. The temperature is not well constrained from HCO$^+$, but compatible with that found
from CO isotopologues. Overall, the non-Keplerian motion detected in Paper II seems less obvious in the
HCO$^+$ data: the best fit inclination is slightly lower, giving a rotation velocity more in agreement with
the stellar mass. Since our low angular resolution data masks out HCO$^+$ emission from inner, highly
inhomogeneous  disk regions, the corresponding velocity profile seems to be closer to a
Keplerian law. However, if the inner radius is fixed to 75 AU, the velocity index derived from HCO$^+$ 
becomes $0.36 \pm 0.04$, in agreement with Paper II.

We use derived disk parameters (Col.~2 in Table~\ref{initial_params}) to determine
column densities for all observed molecules. The derived column density of HCO$^+$
and upper limits for other species are summarized in Table~\ref{chi2_fit}. These column densities scale linearly with
the assumed temperature, but are independent from any of the other parameters.

\section{Modeling \label{modelling}}
As a next step, we try to reproduce the measured column densities and the upper limits altogether using
an advanced disk physical model and the chemical structure from a robust chemical model, similar to
Paper I. The overall iterative fitting is easier than in Paper~I since we refrain from comparing deconvolved
observed and simulated spectral maps and base our analysis on the observed quantities derived with
the $\chi^2$-minimization method in the {\it uv}-plane. This allows us to exclude computationally expensive
radiative transfer modeling with beam convolution from consideration.

To simulate the disk physical structure we utilize
a passive flared 2D disk model of \citet{DD03} with a vertical temperature gradient  and non-grey dust opacities.
Some input disk and stellar parameters are taken as determined by the $\chi^2$-fitting
or found in previous works \citep{vdAnc98,Thi04,Pie05,Sem05}, see Table~\ref{initial_params}.
Other key parameters, like
the radial slope of surface density, the disk mass, and the cosmic ray ionization rate, are found by iterative
fitting of the observed data (Table~\ref{best_fit_params}).

\begin{table}[th]
\caption{List of  disk and stellar
parameters using the $\chi^2$ minimization in the UV plane}
\label{initial_params}
\centering
\begin{tabular}{l l l}
\hline\hline  

& \multicolumn{2}{l}{disk parameters derived from}  \\
& $^{13}$CO (Paper II) &    HCO$^+$  (this paper) \\

\hline

V$_\mathrm{LSR}$      (km\,s$^{-1}$)    &   5.87 & 5.87 $\pm 0.03$ \\
Inclination ($^\circ$) & 35 & 21 $\pm 16$ \\
V sin(i) (km\,s$^{-1}$) & 1.67 & 1.49 $\pm 0.07 $ \\
Velocity exponent & 0.40 &  0.40 $\pm 0.04$ \\
Position angle ($^\circ$) & -30 &   -20 $\pm 4$\\
Inner radius  (AU) & 75 & 110 $\pm 16$ \\
Outer radius  (AU) & 1000 & $> 800$ \\
Temperature (K) & 35 & [35] \\
~~~exponent q   & 0.1 &  [0.1]   \\
Column density (cm$^{-2}$) &  6.2\,$\times$\,$10^{22}$  & 6.2 $\pm 1.2\,10^{12}$\\
~~~exponent & 2.5 & 3.5 $\pm 0.5$ \\
Line width      (km\,s$^{-1}$)    & 0.40 & 0.23 $\pm 0.10$ \\
 \hline
\end{tabular}\\

\smallskip
The fixed parameters are taken from Table I of Paper II, as an average of
the results from the $^{13}$CO.  The column density is for H$_2$ (Col 2) and HCO$^+$ (Col 3) at 250 AU.
\end{table}     

\tabcolsep3.2mm
\begin{table}[h]
\caption{List of disk and stellar parameters used in the best-fit model}
\label{best_fit_params}
\centering
\begin{tabular}{llll}
\hline
\hline
Parameter         & Dimension      & Fixed           & Derived \\
\hline
Stellar radius    & $R_{\sun}$     & 2.5             & ... \\
Stellar mass      & $M_{\sun}$     & 2.4            & ... \\
Effective temperature      & K     & 10\,000        & ... \\
Stellar UV RF     & $\chi_0$       & 100\,000       & ... \\
at 100 AU         & & & \\
CRP ionization rate & s$^{-1}$      & ...             & $4\,10^{-18}$ \\
Inner disk radius &   AU           & 70               & ... \\
Outer disk radius &   AU           & 1\,100           & ... \\
Disk mass         &  $M_{\sun}$    &    ...           & 0.02     \\
Surface density                    &                   &         \\
profile           &  ...           &     ...           & -2.15        \\
Grain size        &  $\mu$m        &    0.12          & ...  \\
Dust-to-gas       &  ...           &    0.01         & ...  \\
mass ratio        & & & \\
 \hline
\end{tabular}\\

\smallskip
The fixed stellar and disk parameters are taken from \citet{Pie05}.
\end{table}

We assume that the central A0e star has an effective temperature of 10\,000~K, a radius of 2.5$R_{\sun}$,
and a mass of 2.4 solar masses \citep{vdAnc98,vdAnc00}. The dust grains are modeled as compact spheres of
uniform 0.12~$\mu$m radius made
of pure amorphous silicates with the optical data taken from \citet{DL84}. The standard $1\%$ dust-to-gas
mass ratio is used. Utilizing the results of Papers~I and II, the disk inner radius is assumed to be
$\sim 70$~AU, the outer radius is 1\,100~AU, and the disk age is between 2 and 5~Myr.

The disk is illuminated by the UV radiation from the central
star and by the interstellar UV radiation. The intensity of the un-attenuated stellar UV flux is calculated
using the \citet{kurucz} ATLAS9 of stellar spectra and converted
to the $\chi_*=10^5 \chi$ factor at 100~AU, where the  factor $\chi$ is
the mean interstellar UV field of \citet{G}. The UV intensity at a given disk
location is calculated as a sum of the stellar and interstellar components that are scaled down by the
visual extinction in vertical direction and in direction to the central star
(1D plane-parallel approximation). We model the attenuation of cosmic rays (CRP) by Eq.~(3) from
\citet{Red2} and vary the initial value of the ionization rate $\zeta_\mathrm{CRP}$ between about
$10^{-18}$ and $10^{-17}$~s$^{-1}$ to match the HCO$^+$ data.
In the disk interior, ionization due to the decay of short-living radionuclides is taken into account,
assuming an ionization rate of $6.5\cdot10^{-19}$~s$^{-1}$ \citep{Finocchi_Gail97}.
The stellar X-ray radiation is assumed to be weak in intermediate-mass stars due
to a lack of the dynamo mechanism and thus is neglected.
Though the observed X-ray luminosity of the AB Aur is non-negligible, its spectrum is dominated by
soft photons with energies below $\sim 1$~keV \citep{Telleschi_ea07}, which cannot penetrate easily deep
into the disk.

The gas-grain time-dependent chemical model
adopted in this study is mostly the same as in Paper~I,
but with two major modifications. First, we now use
osu.2007, the latest version of the Ohio State University (OSU)
database of gas-phase reactions\footnote{See: {\it https://www.physics.ohio-state.edu/~eric/research.html}}
\citep{osu}, with all recent updates to the reaction rates. 
Second, we utilize a standard rate approach to the surface chemistry modeling but without H and H$_2$
quantum tunneling \citep{Katz_ea99}. So, only thermal hopping of surface reactants is allowed.
The surface reactions together with desorption energies are taken from
\cite{Garrod_Herbst06}, which is mostly based on previous studies of
\cite{Hasegawa_ea92} and \cite{Hasegawa_Herbst93}. 
Finally, rates of several tens of photodissociation and photoionization
reactions are updated according to
\citet{vDea_06}, using the UV spectral shape typical of a Herbig Ae star and the ISM-like
dust grain optical properties.  

Overall, the disk chemical network consists of about 650 species made of 13 elements and 7300 reactions. Using
this time-dependent chemical model,
distributions of the molecular abundances and column densities for the
considered species are simulated over 5~Myr of evolution.

\section{Results and Discussion \label{results}}
\begin{figure}[th]
\includegraphics[width=0.242\textwidth,clip=1]{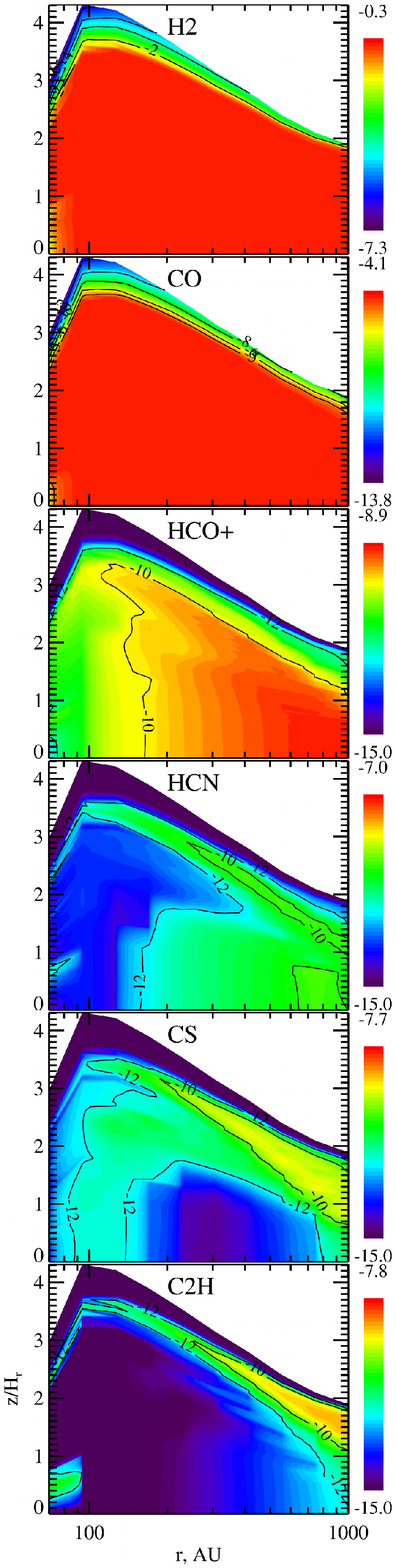}    
\includegraphics[width=0.237\textwidth,clip=1]{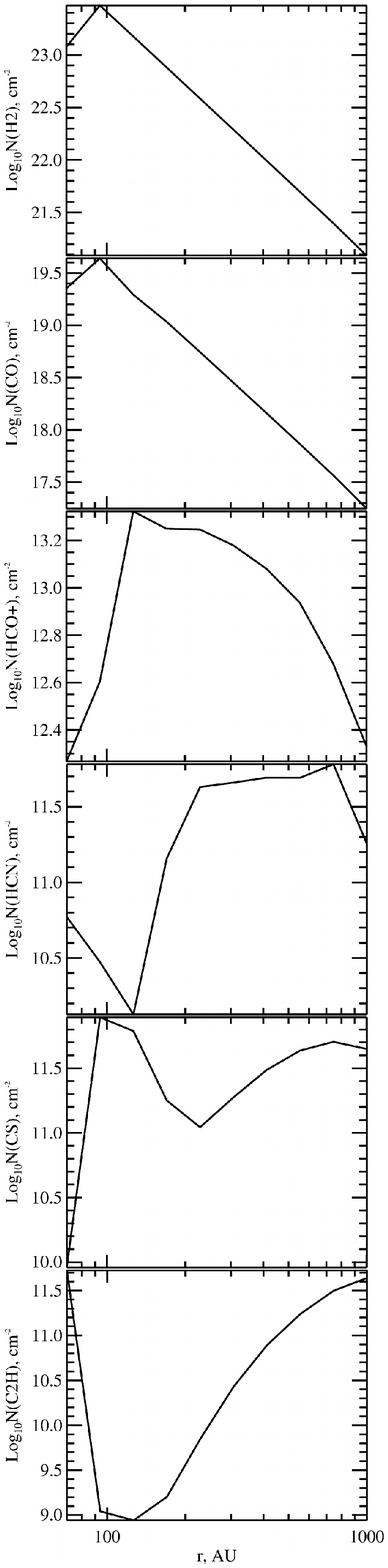}    
\caption{(Left) Modeled abundance distributions in the disk of AB Aur at 2~Myr (relative to
total amount of hydrogen nuclei) for H$_2$, CO, HCO$^+$, HCN, CS, and C$_2$H. The vertical axis is
angle in radians.
(Right) The corresponding radial distributions of the column densities (cm$^{-2}$).}
\label{dima}
\end{figure}

The Fig.~\ref{dima} shows the calculated 2D abundance distribution
for all observed molecular species. The vertical scale is given in units of
pressure scale height calculated for the midplane temperature
\citep{DDG03}. The corresponding vertically integrated column densities versus the radius are shown
and also listed in Table~\ref{chi2_fit}. This best-fit physico-chemical model
of the AB Aur disk is obtained after about 20 iterations by varying the cosmic ray ionization rate, total disk mass,
and surface density exponent. All values  refer to a radius of 250 AU.
Upper limits for CS and \choh\ were derived using the 2\,$\sigma$ errors of the $\chi^2$-minimization.
For these estimates, the H$_2$ column density of $N$(H$_2$) = 6.2\,$\times$\,$10^{22}$ cm$^{-2}$ was
adopted from Paper II as a reference. We have adopted an $^{13}$C/$^{12}$C isotopic ratio of 70.

In agreement with previous studies and the observations of T Tauri
disks, chemical stratification is apparent in the disk around AB Aur
\citep[see Fig.~5 in][]{Dut07}. The disk surrounding the central A0e star is hotter and has a
warm midplane  ($T \ga 20$~K) and, therefore, CO molecules do not freeze out while other molecules do.
Due to self-shielding, energetic UV radiation from AB Aur cannot dissociate a large fraction of CO.
This leads to a high CO column density of about $3\,10^{18}$~cm$^{-2}$ at 250~AU, with a relative CO
abundance of 7\,10$^{-5}$. It seems that such a
low surface CO population in the disk of AB Aur may lead to inhibited catalytic formation
of complex molecules like methanol via hydrogen addition reactions on dust surfaces.
In contrast, in the DM Tau disk a substantial fraction of CO resides on dust grains, so that the averaged
relative CO abundance in the gas phase is 6 times lower, $\sim 10^{-6}$.
We found that, within adopted disk masses of $0.001-0.1M_{\rm sun}$, the radial profile of the CO column
density obtained by chemical modeling strictly follows the input surface density profile with a scaling factor.

\tabcolsep3mm
\begin{table*}[th]
\caption{Observed and modeled column densities $N$ in AB Aur and DM Tau.}
\label{chi2_fit}
\centering
\begin{tabular}{cccccccccccl}
\hline\hline  

Molecule   &\ \ \   & \multicolumn{3}{c}{$\chi^2$-minimization  method} &\ \ \ &  \multicolumn{2}{c}{Chemical model} & DM Tau \\
   &&  N   & 1 $\sigma$   &  N/N($^{13}$CO)$^{\rm (1*)}$ &&  N  &  N/N($^{13}$CO)$^{\rm (2*)}$ & N/N($^{13}$CO)$^{\rm (1*)}$\\
          & & [\cmm] & error & &&   [\cmm]      & & \\
\hline
H$_2$              && $6\,10^{22}$  &  $1\,10^{22}$ & $1.5\,10^{6}$    && $5\,10^{22}$  & $1.3\,10^{6}$     & $1\,10^{7}$  \\
$^{13}$CO$^{(*3)}$&& $4\,10^{16}$  &  $5\,10^{15}$ &  1               && $4\,10^{16}$  &  1                 & 1\\
\hco               && $6\,10^{12}$  &  $3\,10^{11}$ & $1.5\,10^{-4}$  && $1.5\,10^{13}$ &  $4\,10^{-4}$     & $2\,10^{-3}$\\
HCN                && $5\,10^{11}$  &  $3\,10^{11}$ & $1.3\,10^{-5}$  && $4\,10^{11}$   &  $10^{-5}$         & $7\,10^{-4}$\\
CS                 && $3\,10^{12}$  &  $3\,10^{12}$ & $< 8\,10^{-5}$  && $2\,10^{11}$   &  $5\,10^{-6}$      & $3\,10^{-4}$\\
\cch               && $2\,10^{13}$  &  $2\,10^{13}$ & $< 5\,10^{-4}$  && $10^{10}$      &  $2.5\,10^{-7}$    & $10^{-3}$\\
\choh              &&   0           &  $7\,10^{15}$ & $< 2\,10^{-1}$   &&  0             &  0                  &  0 \\
\hline
\end{tabular}

\smallskip
$^{\rm (1*)}$  Relative to the $^{13}$CO column density at 250~AU obtained by the $\chi^2$-minimization method,\\
$^{\rm (2*)}$  Relative to the $^{13}$CO column density at 250~AU obtained by the chemical modeling,\\
$^{\rm (3*)}$  see results reported by Pietu et al.\ (2005).
\end{table*}   

The chemically related molecular ion, HCO$^+$, is directly produced from CO by ion-molecule reactions
with H$_3^+$ and gets destroyed by dissociative recombination. In turn, formation of H$_3^+$ is
solely due to interactions
of cosmic ray particles with molecular hydrogen \citep[see e.g.,][]{Oka_2006}. Thus, the HCO$^+$ column density
is sensitive to adopted cosmic ray ionization rate and, to some extent, temperature structure. Our best-fit model
reproduces the observed HCO$^+$ column density at 250~AU, $6\,10^{12}$~cm$^{-2}$, within a factor of 2.5
($N_{mod}({\rm HCO}^+) \approx 10^{13}$~cm$^{-2}$), which is comparable to observational and inherent chemical
uncertainties \citep{uncr}.

In contrast to the DM~Tau
disk where HCO$^+$ is mostly concentrated in the intermediate layer due to freeze out of CO, HCO$^+$ reaches the highest
concentration in the midplane of the AB Aur disk. The HCO$^+$ ion was found to be very sensitive to the adopted value
of the CRP flux and radial density distribution. The acceptable fit to observations was only possible when
the CRP ionization
rate was lowered by a factor of 3--5 compared to the standard value of $1.3\,10^{-17}$~s$^{-1}$, the surface
density profile
is $\approx -2.15$, and the disk mass is between $0.008-0.012M_{\rm sun}$. With lower disk masses, the predicted
$^{13}$CO column
density was lower than the observed value ($4\,10^{16}$~cm$^{-2}$ at 250~AU), while with other density profiles it was
impossible to get proper CS, C$_2$H and HCN column densities. The model results dependent on the adopted
inner radius -- with radii smaller than about 50~AU and fixed total disk mass the modeled  HCO$^+$ and
CO concentrations were lower than the observed values.
As we used the newly measured inner radius of $\sim 70$~AU (see paper~II),
it is not surprising that the modeled distribution  of
HCO$^+$ in the AB Aur disk differs from the results of Paper~I (see Fig.~5 therein).

The reason why the CRP ionization rate could be lower in AB Aur than a widely accepted value
($\approx 10^{-17}$~s$^{-1}$) is manyfold. First of all, the standard ionization rates of He and H$_2$ can be
uncertain by
a factor of 2 \citep{Wakelam_ea06}. Second, if a self-generation of magnetic fields in protoplanetary disks
by a dynamo-like mechanism takes place \citep{DG_94}, then the
cosmic ray particles will be scattered and will not reach disk midplanes -- exactly the zone where a
significant fraction of HCO$^+$ exists in the AB Aur disk. Third, the AB Aur disk has a complex and clumpy structure
\citep{Fuk04, Pie05, Lin06}, and, consequently, efficiency of dissociative recombination of HCO$^+$ will be different in
clumps compared to an inter-clump medium. Our axisymmetric disk model does not take this effect into account.
Last, there may be some missed important reaction pathways to HCO$^+$ or improperly derived reactions rates.

Abundances and column densities of other observed molecules are also explained by this model.
The chemically active C$_2$H radical is only abundant in dilute, irradiated disk surface and quickly
converted into heavier, more complex species in the denser region adjacent to the midplane (Fig.~\ref{dima}).
On the other hand, HCN and CS are more widely distributed across the AB Aur disk, though both these molecules
are underabundant in the midplane where they stick to grains and surface processed.
Note that the column densities of CO and thus HCO$^+$ decrease with radius, whereas the column densities of
C$_2$H and HCN slightly increase with radius. The CS column density stays nearly constant for large radii,
$r\ga 200$~AU, and is found to be stable against  iterative variations of the input parameters. It is HCN that
depends on the adopted value of the surface density exponent. As this molecule can be easily photodissociated
by stellar UV photons, its abundance is sensitive to details of the
disk vertical structure, in particular, shielding by dust.
Presumably, all modeled column densities agree well with the observed values or upper limits due to the
lack of sensitivity of the measurements. On the other hand, the derived radial profiles are
shallower than extracted from observations by the $\chi^2$-minimization analysis. The
surface density is a measure of the total disk mass, which is usually obtained from thermal dust emission and poorly
known dust properties (gas-to-dust ratio, opacities), and thus can be rather uncertain (see discussion in Paper~I).

It is interesting to compare the molecular contents of the AB Aur disk and of another well-studied
disk around the  T Tauri star DM Tau. The DM Tau disk has a similar size of $\sim 1\,000$~AU 
with a hole of a few AU in size \citep{Cal05}, 
is more evolved ($\sim 5$~Myr), and more massive ($\sim 0.05\,M_{\sun}$) \citep[see Table~2 in][]{Dut07}.
In Table~\ref{chi2_fit} the observed column densities rescaled to the $^{13}$CO value at 250~AU are listed
for the DM Tau disk as well (last column). These rescaled values are less sensitive to possible errors in the
estimate of the disk mass and surface density.

It is apparent that relative column densities for all observed species are lower in the disk of AB Aur compared to the
values observed in the DM Tau disk. Given that the absolute column density of $^{13}$CO at 250~AU is similar in
both disks, it implies a poor molecular content of the AB Aur disk (except of CO and H$_2$).
However, the disk around DM Tau is about 4 times
more massive than the AB Aur disk. The fact that both 
disks harbor similar amounts of CO despite significant difference
in their masses is likely a direct manifestation of severe freeze out in the cold midplane of the DM Tau source and/or
difference in grain properties.

The lower mass of the hotter AB Aur disk is likely the reason for its molecule-deficient
gas content. Since the total dust shielding is lower, while the UV luminosity of the Herbig A0 star is orders
of magnitude higher than that of DM Tau, the photodissociation of many molecules is more effective in the disk
around AB Aur. The age difference between the two sources is a negligible factor for the results of chemical
modeling, even though gas-grain models usually do not reach a steady-state within a few Myr. The same is true
for surface chemistry, which is not much of importance for most of the observed species, apart from methanol.

\section{Conclusions and Summary}
The AB Aur environment is clearly dominated by the envelope, as shown by the detections made with the 30-m
telescope and the lack of detections with the IRAM array in short (1-3 hours) integrations.
The disk has an inner hole in the gas distribution with a radius of $\approx$ 70 AU.
Using a gas-grain chemical model with surface reactions coupled to
a flared passive disk model, we reproduce all observed column densities and upper limits.
Note that such a comparison between the advanced chemical model and the interferometric
observations are hampered by the lack of knowledge on the H$_2$ surface density. The surface density
of the AB Aur disk is derived from the millimeter dust emission, a procedure prone to
uncertainties. Part of the discrepancies
between the model and the observations may also be due to inhomogeneous structure of the AB Aur
disk and/or grain evolution in central disk
regions that is not considered in the calculations. Modeled and observed column densities relative to $^{13}$CO are
both lower for the AB Aur disk than the values measured in DM Tau. The absolute amounts of CO gas are similar
 in  both disks. The poor molecular content of the AB Aur system can be explained
by much more intense UV irradiation by the central A0 star and less effective dust shielding in the low-mass disk
compared to the more massive disk around the less luminous M1 star DM Tau. The indirectly inferred cosmic ray
ionization rate that is needed to fit the HCO$^+$ data is a factor of 2--5 lower than the standard value, which can
be explained by scattering of CRPs in a magnetized AB Aur disk or a clumpy disk structure.

\begin{acknowledgements}
We acknowledge all the Plateau de Bure IRAM staff for their help
during the observations. SG, AD, VW, FH, and VP are financially
supported by the French Program ``Physique Chimie du Milieu Interstellaire'' (PCMI).
\end{acknowledgements}


\end{document}